\documentclass[twocolumn,floatfix,showpacs,amssymb,amsmath,pra,aps,10pt]{revtex4-1}
\usepackage{graphicx}
\usepackage{amsmath}

\begin{document}

\def\la{\langle}
\def\ra{\rangle}

\title{Precision frequency measurements with interferometric weak values}
\author{David J. Starling$^\star$, P. Ben Dixon, Andrew N. Jordan and John C. Howell}
\affiliation{Department of Physics and Astronomy, University of Rochester, Rochester, New York 14627, USA}

\begin{abstract}
We demonstrate an experiment which utilizes a Sagnac interferometer to measure a change in optical frequency of 129 $\pm$ 7 $\mathbf{\mbox{kHz}/\sqrt{\mbox{Hz}}}$ with only 2 mW of continuous wave, single mode input power. We describe the measurement of a weak value and show how even higher frequency sensitivities may be obtained over a bandwidth of several nanometers. This technique has many possible applications, such as precision relative frequency measurements and laser locking without the use of atomic lines.
\end{abstract}


\maketitle

\section{Introduction}

Precision frequency measurements \cite{Junttila1990,Ishikawa1993,Jones2000} of a stabilized laser source are of great importance in the field of metrology \cite{Udem2002} as well as atomic, molecular \cite{Falke2008} and optical physics \cite{Eisele2009}. Here we show that weak values \cite{Aharonov1988,Duck1989,Dressel2010} in an optical deflection measurement experiment \cite{Dixon2009} can produce frequency shift resolutions down to 129 $\pm$ 7 $\mathbf{\mbox{kHz}/\sqrt{\mbox{Hz}}}$ with only 2 mW of continuous wave optical power. By performing a weak measurement of the deflection of an infrared laser source that has passed through a weakly dispersive prism, we are able to measure a change in optical frequency comparable to precision Fabry-Perot interferometers \cite{Hori1989,Nishimiya2004,Notcutt2005}. This technique is relatively simple, requiring only a few common optical components and operating at atmospheric pressure. Additionally, we show that this technique has low noise over a large range of response frequencies, making it desirable for many applications such as Doppler anemometry \cite{Tropea1995}, tests of the isotropy of light propagation \cite{Eisele2009} or laser locking without the use of high finesse Fabry-Perot interferometers \cite{Pohl2010} or atomic lines.

First developed as a way to understand preselected and postselected quantum measurements and how they relate to time-reversal symmetry in quantum mechanics, the \emph{weak value} $A_w$ of an operator ${\bf A}$ was introduced in a seminal 1988 paper by Aharonov, Albert and Vaidman (AAV) \cite{Aharonov1988}. The weak value is given by $A_w = \la \psi_f | {\bf A} |\psi_i\ra/\la \psi_f  |\psi_i\ra$, where $\{|\psi_{i,f}\ra\}$ are the preselected and postselected states of the system, respectively. This quantity, which is likened to the expectation value of ${\bf A}$, can have seemingly strange behavior, particularly in the limit where the preselected and postselected states are nearly orthogonal. While numerous experiments have validated the initial claims of the AAV paper \cite{Ritchie1990,Pryde2005,Dixon2009,Starling2010}, there are new developments concerning the interpretation of preselected and postselected weak measurements \cite{Duck1989,Aharonov1991,Steinberg2010,Dressel2010}.

Weak values are a result of a so-called weak measurement, \emph{i.e.,} a measurement which gains only partial information about the state of a system. Unlike von Neumann measurements, a weak measurement disturbs the measured state of the system only minimally. For this reason, weak measurements have been useful in reconsidering Hardy's paradox \cite{Hardy1992,Lundeen2009} as well as making meaningful, sequential measurements of noncommuting observables \cite{Mitchison2007}. Furthermore, due to the denominator of $A_w$, there can be a large amplification of the weak value when the preselected and postselected states are nearly orthogonal; as a result, there have been a number of experimental results published in the field of optical metrology \cite{Hosten2008,Dixon2009,Starling2009,Starling2010}. There is also a vast array of results, both theoretical \cite{Williams2008,Howell2010,Brunner2010,Dressel2010} and experimental \cite{Ritchie1990,Pryde2005}, which have gone a long way to further our understanding of the weak measurement process.

\begin{figure}
\includegraphics[scale=1]{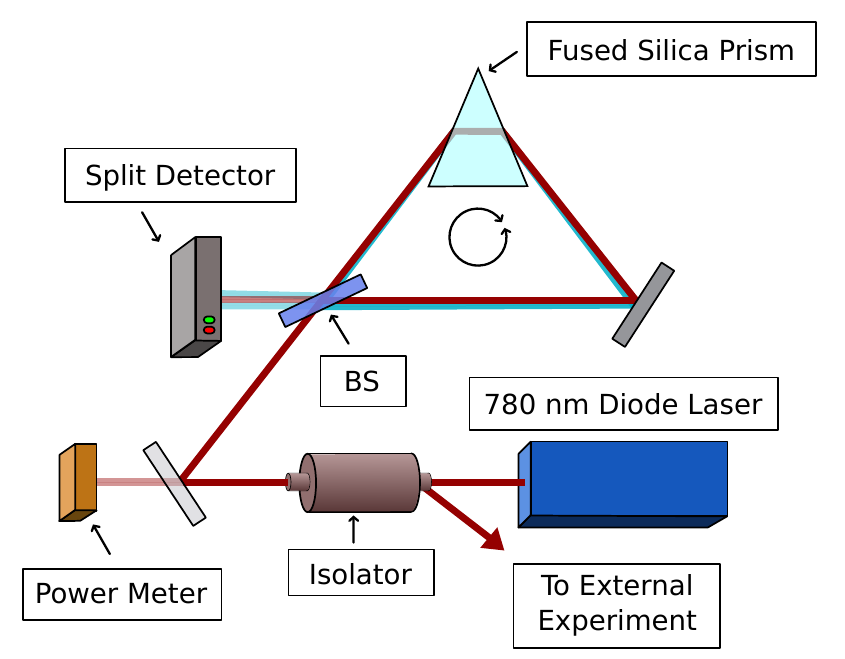}
\caption{(Color online) A Gaussian laser beam passes through a Sagnac interferometer consisting of a 50:50 beam splitter (BS), a mirror and a prism. The prism weakly perturbs the direction of the beam as the frequency of the laser source is modulated, denoted by the red and blue beam paths. We monitor the position of the light entering the dark port of the interferometer. We lock the input power to the interferometer using a power measurement before the BS. The majority of the light exits the interferometer via the bright port and is collected with an isolator for use in an experiment.}
\end{figure}

\section{Theory}

We describe here the frequency amplification experiment shown in Fig.\ 1 by further developing the ideas of Ref.\ \onlinecite{Dixon2009}. Although the actual experiment uses a classical beam, we choose to characterize the weak value effect one photon at a time; this is valid, owing to the fact that we consider here a linear system with a coherent laser beam modeled as a linear superposition of Fock states \cite{Howell2010}. 

In this experiment, a single-mode Gaussian beam of frequency $\omega$ and radius $\sigma$ passes through an optical isolator, resulting in linearly polarized light. We assume that the radius is large enough to ignore divergence due to propagation. Light then enters a Sagnac interferometer containing a 50:50 beam splitter (BS), a mirror and a prism. The beam travels clockwise and counter-clockwise through the interferometer, denoted by the system states given by \{$|\!\circlearrowright\ra, |\!\circlearrowleft\ra$\}; we write the photon meter states in the position basis as \{$|x\ra$\}, where $x$ denotes the transverse, \emph{horizontal} direction. 

Initially, the interferometer (including the prism) is aligned such that the split photon wave function spatially overlaps (\emph{i.e.,} the photons travel the same path whether by $|\!\circlearrowright\ra$ or $|\!\circlearrowleft\ra$). After the interferometer is aligned, the photons traversing each path receive a small, constant momentum kick in the \emph{vertical} direction; this vertical kick is controlled by the interferometer mirror and results in a misalignment. Due to its spatial asymmetry about the input BS, this momentum kick creates an overall phase difference $\phi$ between the two paths. By adjusting the interferometer mirror, we can control the amount of light that exits the interferometer into the dark port. While the amplified signal ultimately depends on the value of $\phi$, and therefore on the magnitude of the misalignment, the \emph{signal to noise ratio} (SNR) is unaffected (discussed below). 

We then let $k(\omega)$ represent the small momentum kick given by the prism to the beam (after alignment) in the horizontal $x$-direction. The \emph{system} and \emph{meter} are entangled via an impulsive interaction Hamiltonian \cite{Dixon2009} (resulting in a new state $|\psi_i\ra\rightarrow |\Psi\ra$) such that a measurement of the horizontal position of the photon after it exits the interferometer gives us some information about which path the photon took. 

We consider a horizontal deflection that is significantly smaller than the spread of the wave packet we are trying to measure, \emph{i.e.,} $k(\omega) \sigma \ll 1$. In this approximation, we find that the postselected state of the photons exiting the dark port is given by 
\begin{equation}
\la \psi_f |\Psi \ra= \la \psi_f |\psi_i\ra \int dx \psi(x) |x\ra  \exp[-i x A_w k(\omega)],
\label{state}
\end{equation}
where the weak value, defined above, is given by $A_w  = -i\cot(\phi/2) \approx -2i/\phi$ for small $\phi$.

\begin{figure}
\includegraphics{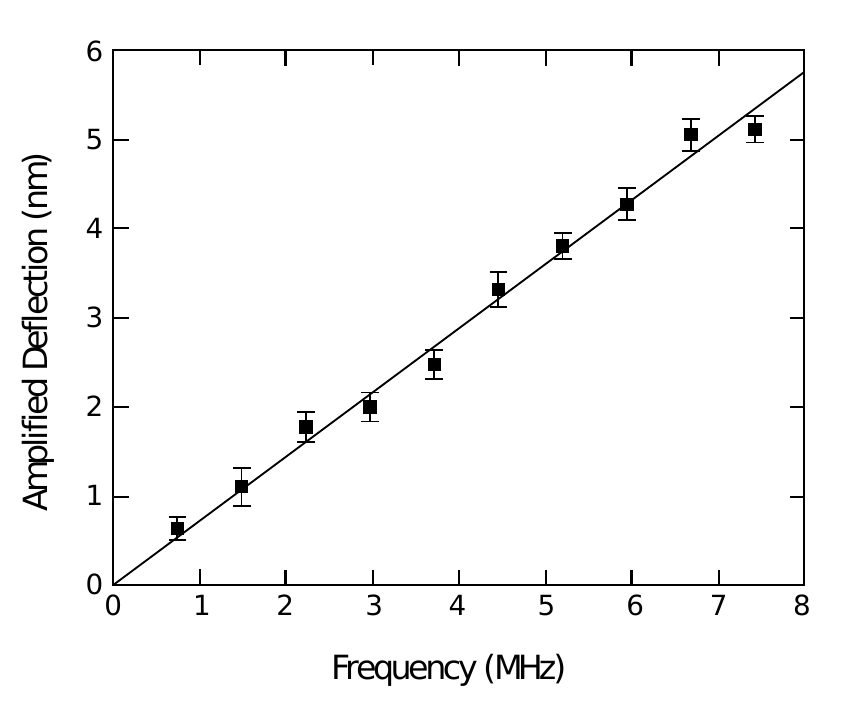}
\caption{The position of the postselected beam profile is measured as we modulate the input laser frequency of the interferometer. The modulation oscillates as a sine wave at 10 Hz and the signal from the split detector is frequency filtered and amplified. The error bars are given by the standard deviation of the mean. The minimum frequency change measured here is around 743 kHz with an effective integration time of 30 ms. The weak value amplification is approximately 79.}
\end{figure}

There are two interesting features of Eq.\ (\ref{state}). First, the probability of detecting a photon has been reduced to $P_{ps} =|\la \psi_f  |\psi_i\ra|^2  = \sin^2 (\phi/2)$, and yet the SNR of an ensemble of measurements is nearly quantum limited \cite{Starling2009} despite not measuring the vast majority of the light. Second, the weak value (which can be arbitrarily large in theory) appears to amplify the momentum kick $k(\omega)$ given by the prism; the resulting average position is given by $\la x \ra _W= 2 k(\omega) \sigma^2 |A_w| \approx 4 k(\omega)  \sigma^2 /\phi$, where the angular brackets denote an expectation value. We can compare this to the standard deflection caused by a prism measured at a distance $l$ which is given by $\la x \ra \approx l k(\omega)/k_0$, where $k_0$ is the wavenumber of the light.

In order to predict the deflections $\la x \ra$ or $\la x \ra_W$, we must know the form of $k(\omega)$. For a prism oriented such that it imparts the minimum deviation on a beam, the total angular deviation is given by $\theta(\omega) = 2 \sin^{-1}\left[n(\omega) \sin(\gamma/2)\right]-\gamma$, where $n(\omega)$ is the index of refraction of the material and $\gamma$ is the angle at the apex of the prism \cite{Pedrotti1993}. However, we are only interested in the small, frequency-dependent angular deflection $\delta(\omega) = \Delta\theta = 2 \,\Delta n \{[\sin(\gamma/2)]^{-2}-[n(\omega)]^2\}^{-1/2}$, where $\Delta n$ ($\Delta \theta$) is the index change in the prism (angular deflection of the beam) for a given frequency change of the laser. The small \emph{momentum kick} is expressed as $k(\omega)=\delta(\omega) k_0$. We can then write the amplified deflection as
\begin{equation}
\la x \ra_W \approx \frac{8 k_0 \sigma^2 (\Delta n/\phi)}{\sqrt{[\sin(\gamma/2)]^{-2}-[n(\omega)]^2}}.
\label{deflection}
\end{equation}

The frequency-dependent index $n(\omega)$ of fused silica, which was used in this experiment, can be modeled using the Sellmeier equation \cite{Malitson1965}. We can therefore calculate the expected $\la x \ra_W$ using Eq.\ (\ref{deflection}). However, to compute the ultimate sensitivity of this weak value frequency measurement, we must include possible noise sources. If we consider only shot-noise from the laser, the SNR for small $\phi$ is approximated by 
\begin{equation}
\mathcal{R} \approx \sqrt{\frac{8 N}{\pi}} k_0 \sigma \delta(\omega),
\label{snr}
\end{equation}
as shown in Ref.\ \onlinecite{Starling2009}, where $N$ is the number of photons used in the interferometer. Note that $N$ is not the number of photons striking the detector, which is given by $N P_{ps}$. By setting $\mathcal{R}=1$ and using modest values for $N$, $\sigma$ and $\omega$, we find that frequency sensitivities well below 1 kHz are possible. However, other sources of noise, such as detector dark current, radiation pressure and environmental perturbations will reduce the sensitivity of the device.

\section{Experiment}

In our experimental setup (shown in Fig.\ 1), we used a fiber-coupled 780 nm external cavity diode laser with a beam radius of $\sigma=388\,\mu$m. The frequency of the laser was modulated with a 10 Hz sine wave using piezo controlled grating feedback. The frequency control was calibrated using saturation absorption spectroscopy of the hyperfine excited states of the rubidium D2 line\cite{Steck2010}: $F=3 \rightarrow$ $F'=\{2-4\mbox{ crossover, } 3-4 \mbox{ crossover, } 4\}$ transitions of rubidium 85 and the $F=2 \rightarrow$ $F'=\{1-3\mbox{ crossover, } 2-3 \mbox{ crossover, } 3\}$ transitions of rubidium 87. Linearly polarized light  was divided before the interferometer using a 50:50 BS (although an imbalanced ratio here would be ideal for practical applications). The light in one port was measured with a photodiode and used to lock the power at 2 mW with an acousto-optic modulator before the fiber. The interferometer was approximately $l = 27$ cm in length; the mirror used to adjust $\phi$ was approximately 6 cm from the input BS (measured counter clockwise) and the prism, made of fused silica, was approximately 5 cm from the input BS (measured clockwise). Although the prism was not symmetrically placed in the interferometer as described in the theory above, the results are the same aside from a global offset in position which can be subtracted off during processing. The interferometer was first aligned to minimize light in the dark port and then, using the aforementioned mirror, misaligned to allow a small percentage of the light ($\sim$2-5\%) into the dark port. The position of this light was measured using a split detector (New Focus model 2921). The signal was passed through two 6 dB/octave bandpass filters centered at 10 Hz and amplified by a factor of about $10^4$. 

For Fig.\ 2, we measured the peak of the deflection in each 100 ms cycle, repeated 25 times;  we computed the average and the standard deviation of this set as we varied the change in the optical frequency. We find that the amplified deflection is a linear function of oscillating optical frequency given by about 720 $\pm$ 11 pm/MHz. Compared to the unamplified deflection of about 9.1 pm/MHz given by the expression for $\la x \ra$, this gives an amplification factor of 79 $\pm$ 1.2 and a computed $P_{ps}$ of 1.3\%; this agrees with the measured $P_{ps}$ of 2-5\% if we include the extra light present in the signal due to phase front distortions from imperfect optics.

\begin{figure}
\includegraphics[scale=1]{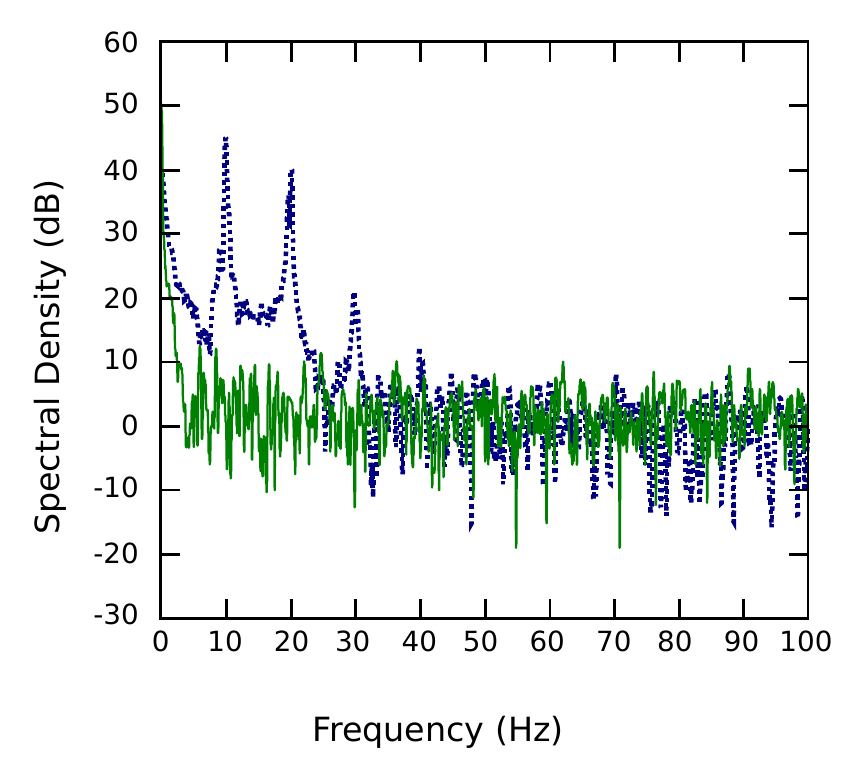}
\caption{(Color online) We show the noise spectrum for a passive system (green, solid trace) and for a driven system (blue, dashed trace), where the laser frequency modulation is 7.4 MHz at 10 Hz. We see that the first harmonic of the signal is about 5 dB down from the fundamental and the third harmonic is nearly 25 dB down. For a 7.4 MHz change in laser frequency, we see that the noise is approximately 35 dB below the signal, demonstrating the low-noise nature of this measurement.}
\end{figure}

A characteristic noise scan was taken and plotted in Fig.\ 3 with and without frequency modulation. The signal was passed directly from the split detector into the oscilloscope before performing a fast Fourier transform. Data was taken with and without a 7.4 MHz optical frequency modulation to show the noise floor over a large bandwidth. The noise at higher frequencies was similarly flat. Second, to test the range over which this device could function, we optimized the interferometer at the low-frequency end of the laser's tuning range and obtained a SNR of approximately 19 with the 7.4 MHz optical frequency modulation.  We then tuned to the high-frequency end of the laser's tuning range ($\Delta f \approx 141$ GHz), without adjusting or recalibrating the interferometer, and obtained a SNR of 10. In fact, this range can be much larger so long as the weak value condition $k(\omega) \sigma \ll 1$ is satisfied; for our beam radius and optical frequency, we could in principle measure over a range of 5 THz, or about 10 nm. 

For our experimental parameters, we can measure below 1 MHz of frequency change with a SNR around 1, as shown in Fig.\ 2. It should be noted that, although the time between measurements is a full 100 ms, our filtering limits the laser noise to time scales of about 30 ms. For analysis, we take this as our integration time in estimating $N$ for each measurement. The resulting sensitivity for our apparatus is 129 $\pm$ 7 $\mbox{kHz}/\sqrt{\mbox{Hz}}$; \emph{e.g.,} if we had integrated for 1 s instead of 30 ms, this device could measure a 129 kHz shift in frequency with a SNR of 1. The error in frequency comes from the calibration described above. Using Eq.\ (\ref{snr}), we find that the ideal ultimate sensitivity is approximately 67 $\mbox{kHz}/\sqrt{\mbox{Hz}}$. This implies that this apparatus, operating at atmospheric pressure with modest frequency filtering, is less than a factor of two away from the shot-noise limit in sensitivity. This is no longer surprising since we now understand the fact that weak value experiments amplify the signal, but not the technical noise \cite{Hosten2008,Starling2009}.

\section{Conclusion}

With only 2 mW of continuous wave input power, we have measured a frequency shift of 129 $\pm$ 7 $\mbox{kHz}/\sqrt{\mbox{Hz}}$; we have shown that the system is stable over our maximum tuning range of 140 GHz without recalibration and is nearly shot-noise-limited. With more optical power, longer integration and a more dispersive element such as a grating or a prism with $\sin(\gamma/2)n(\omega)\approx1$, the sensitivity of this device can measure frequency shifts lower than 1  kHz, although a higher sensitivity comes at the cost of maximum tuning range. Compare this to commercially available Fabry-Perot interferometers, which report typical resolutions down to 5 MHz and free spectral ranges of only 1-5 GHz. More sensitive Fabry-Perot interferometers exist, yet they require a host of custom equipment to reduce environment noise, including vacuum systems and vibration damping. Moreover, an important advantage of this technique is that a large percentage ($\sim$90\%) of the light used in the interferometer can then be sent off to another experiment (as indicated in Fig.\ 1), allowing for real-time frequency information during data collection. While this device cannot compare to the absolute frequency sensitivity of frequency combs \cite{Jones2000}, we believe that this method is a simple solution for high-resolution, relative frequency metrology and will serve as a valuable laser-locking tool.

\section*{Acknowledgments}

This research was supported by US Army Research Office Grant No. W911NF-09-0-01417,  DARPA Expansion Grant No. N00014-08-1-120 and the University of Rochester.


\begin{thebibliography}{31}%
\makeatletter
\providecommand \@ifxundefined [1]{%
 \@ifx{#1\undefined}
}%
\providecommand \@ifnum [1]{%
 \ifnum #1\expandafter \@firstoftwo
 \else \expandafter \@secondoftwo
 \fi
}%
\providecommand \@ifx [1]{%
 \ifx #1\expandafter \@firstoftwo
 \else \expandafter \@secondoftwo
 \fi
}%
\providecommand \natexlab [1]{#1}%
\providecommand \enquote  [1]{``#1''}%
\providecommand \bibnamefont  [1]{#1}%
\providecommand \bibfnamefont [1]{#1}%
\providecommand \citenamefont [1]{#1}%
\providecommand \href@noop [0]{\@secondoftwo}%
\providecommand \href [0]{\begingroup \@sanitize@url \@href}%
\providecommand \@href[1]{\@@startlink{#1}\@@href}%
\providecommand \@@href[1]{\endgroup#1\@@endlink}%
\providecommand \@sanitize@url [0]{\catcode `\\12\catcode `\$12\catcode
  `\&12\catcode `\#12\catcode `\^12\catcode `\_12\catcode `\%12\relax}%
\providecommand \@@startlink[1]{}%
\providecommand \@@endlink[0]{}%
\providecommand \url  [0]{\begingroup\@sanitize@url \@url }%
\providecommand \@url [1]{\endgroup\@href {#1}{\urlprefix }}%
\providecommand \urlprefix  [0]{URL }%
\providecommand \Eprint [0]{\href }%
\providecommand \doibase [0]{http://dx.doi.org/}%
\providecommand \selectlanguage [0]{\@gobble}%
\providecommand \bibinfo  [0]{\@secondoftwo}%
\providecommand \bibfield  [0]{\@secondoftwo}%
\providecommand \translation [1]{[#1]}%
\providecommand \BibitemOpen [0]{}%
\providecommand \bibitemStop [0]{}%
\providecommand \bibitemNoStop [0]{.\EOS\space}%
\providecommand \EOS [0]{\spacefactor3000\relax}%
\providecommand \BibitemShut  [1]{\csname bibitem#1\endcsname}%
\let\auto@bib@innerbib\@empty
\bibitem [{\citenamefont {Junttila}\ and\ \citenamefont
  {Stahlberg}(1990)}]{Junttila1990}%
  \BibitemOpen
  \bibfield  {author} {\bibinfo {author} {\bibfnamefont {M.-L.}\ \bibnamefont
  {Junttila}}\ and\ \bibinfo {author} {\bibfnamefont {B.}~\bibnamefont
  {Stahlberg}},\ }\href@noop {} {\bibfield  {journal} {\bibinfo  {journal}
  {Appl. Opt.}\ }\textbf {\bibinfo {volume} {29}},\ \bibinfo {pages} {3510}
  (\bibinfo {year} {1990})}\BibitemShut {NoStop}%
\bibitem [{\citenamefont {Ishikawa}\ and\ \citenamefont
  {Watanabe}(1993)}]{Ishikawa1993}%
  \BibitemOpen
  \bibfield  {author} {\bibinfo {author} {\bibfnamefont {J.}~\bibnamefont
  {Ishikawa}}\ and\ \bibinfo {author} {\bibfnamefont {H.}~\bibnamefont
  {Watanabe}},\ }\href {\doibase 10.1109/19.278596} {\bibfield  {journal}
  {\bibinfo  {journal} {IEEE Trans. Instrum. Meas.}\ }\textbf {\bibinfo
  {volume} {42}},\ \bibinfo {pages} {423 } (\bibinfo {year}
  {1993})}\BibitemShut {NoStop}%
\bibitem [{\citenamefont {Jones}\ \emph {et~al.}(2000)\citenamefont {Jones},
  \citenamefont {Diddams}, \citenamefont {Ranka}, \citenamefont {Stentz},
  \citenamefont {Windeler}, \citenamefont {Hall},\ and\ \citenamefont
  {Cundiff}}]{Jones2000}%
  \BibitemOpen
  \bibfield  {author} {\bibinfo {author} {\bibfnamefont {D.}~\bibnamefont
  {Jones}}, \bibinfo {author} {\bibfnamefont {S.}~\bibnamefont {Diddams}},
  \bibinfo {author} {\bibfnamefont {J.}~\bibnamefont {Ranka}}, \bibinfo
  {author} {\bibfnamefont {A.}~\bibnamefont {Stentz}}, \bibinfo {author}
  {\bibfnamefont {R.}~\bibnamefont {Windeler}}, \bibinfo {author}
  {\bibfnamefont {J.}~\bibnamefont {Hall}}, \ and\ \bibinfo {author}
  {\bibfnamefont {S.}~\bibnamefont {Cundiff}},\ }\href@noop {} {\bibfield
  {journal} {\bibinfo  {journal} {Science}\ }\textbf {\bibinfo {volume}
  {288}},\ \bibinfo {pages} {635} (\bibinfo {year} {2000})}\BibitemShut
  {NoStop}%
\bibitem [{\citenamefont {Udem}\ \emph {et~al.}(2002)\citenamefont {Udem},
  \citenamefont {Holzwarth},\ and\ \citenamefont {Hansch}}]{Udem2002}%
  \BibitemOpen
  \bibfield  {author} {\bibinfo {author} {\bibfnamefont {T.}~\bibnamefont
  {Udem}}, \bibinfo {author} {\bibfnamefont {R.}~\bibnamefont {Holzwarth}}, \
  and\ \bibinfo {author} {\bibfnamefont {T.~W.}\ \bibnamefont {Hansch}},\
  }\href {\doibase 10.1038/416233a} {\bibfield  {journal} {\bibinfo  {journal}
  {Nature}\ }\textbf {\bibinfo {volume} {416}},\ \bibinfo {pages} {233}
  (\bibinfo {year} {2002})}\BibitemShut {NoStop}%
\bibitem [{\citenamefont {Falke}\ \emph {et~al.}(2008)\citenamefont {Falke},
  \citenamefont {Kn\"ockel}, \citenamefont {Friebe}, \citenamefont {Riedmann},
  \citenamefont {Tiemann},\ and\ \citenamefont {Lisdat}}]{Falke2008}%
  \BibitemOpen
  \bibfield  {author} {\bibinfo {author} {\bibfnamefont {S.}~\bibnamefont
  {Falke}}, \bibinfo {author} {\bibfnamefont {H.}~\bibnamefont {Kn\"ockel}},
  \bibinfo {author} {\bibfnamefont {J.}~\bibnamefont {Friebe}}, \bibinfo
  {author} {\bibfnamefont {M.}~\bibnamefont {Riedmann}}, \bibinfo {author}
  {\bibfnamefont {E.}~\bibnamefont {Tiemann}}, \ and\ \bibinfo {author}
  {\bibfnamefont {C.}~\bibnamefont {Lisdat}},\ }\href {\doibase
  10.1103/PhysRevA.78.012503} {\bibfield  {journal} {\bibinfo  {journal} {Phys.
  Rev. A}\ }\textbf {\bibinfo {volume} {78}},\ \bibinfo {pages} {012503}
  (\bibinfo {year} {2008})}\BibitemShut {NoStop}%
\bibitem [{\citenamefont {Eisele}\ \emph {et~al.}(2009)\citenamefont {Eisele},
  \citenamefont {Nevsky},\ and\ \citenamefont {Schiller}}]{Eisele2009}%
  \BibitemOpen
  \bibfield  {author} {\bibinfo {author} {\bibfnamefont {C.}~\bibnamefont
  {Eisele}}, \bibinfo {author} {\bibfnamefont {A.~Y.}\ \bibnamefont {Nevsky}},
  \ and\ \bibinfo {author} {\bibfnamefont {S.}~\bibnamefont {Schiller}},\
  }\href {\doibase 10.1103/PhysRevLett.103.090401} {\bibfield  {journal}
  {\bibinfo  {journal} {Phys. Rev. Lett.}\ }\textbf {\bibinfo {volume} {103}},\
  \bibinfo {pages} {090401} (\bibinfo {year} {2009})}\BibitemShut {NoStop}%
\bibitem [{\citenamefont {Aharonov}\ \emph {et~al.}(1988)\citenamefont
  {Aharonov}, \citenamefont {Albert},\ and\ \citenamefont
  {Vaidman}}]{Aharonov1988}%
  \BibitemOpen
  \bibfield  {author} {\bibinfo {author} {\bibfnamefont {Y.}~\bibnamefont
  {Aharonov}}, \bibinfo {author} {\bibfnamefont {D.~Z.}\ \bibnamefont
  {Albert}}, \ and\ \bibinfo {author} {\bibfnamefont {L.}~\bibnamefont
  {Vaidman}},\ }\href {\doibase 10.1103/PhysRevLett.60.1351} {\bibfield
  {journal} {\bibinfo  {journal} {Phys. Rev. Lett.}\ }\textbf {\bibinfo
  {volume} {60}},\ \bibinfo {pages} {1351} (\bibinfo {year}
  {1988})}\BibitemShut {NoStop}%
\bibitem [{\citenamefont {Duck}\ \emph {et~al.}(1989)\citenamefont {Duck},
  \citenamefont {Stevenson},\ and\ \citenamefont {Sudarshan}}]{Duck1989}%
  \BibitemOpen
  \bibfield  {author} {\bibinfo {author} {\bibfnamefont {I.~M.}\ \bibnamefont
  {Duck}}, \bibinfo {author} {\bibfnamefont {P.~M.}\ \bibnamefont {Stevenson}},
  \ and\ \bibinfo {author} {\bibfnamefont {E.~C.~G.}\ \bibnamefont
  {Sudarshan}},\ }\href {\doibase 10.1103/PhysRevD.40.2112} {\bibfield
  {journal} {\bibinfo  {journal} {Phys. Rev. D}\ }\textbf {\bibinfo {volume}
  {40}},\ \bibinfo {pages} {2112} (\bibinfo {year} {1989})}\BibitemShut
  {NoStop}%
\bibitem [{\citenamefont {Dressel}\ \emph {et~al.}(2010)\citenamefont
  {Dressel}, \citenamefont {Agarwal},\ and\ \citenamefont
  {Jordan}}]{Dressel2010}%
  \BibitemOpen
  \bibfield  {author} {\bibinfo {author} {\bibfnamefont {J.}~\bibnamefont
  {Dressel}}, \bibinfo {author} {\bibfnamefont {S.}~\bibnamefont {Agarwal}}, \
  and\ \bibinfo {author} {\bibfnamefont {A.~N.}\ \bibnamefont {Jordan}},\
  }\href {\doibase 10.1103/PhysRevLett.104.240401} {\bibfield  {journal}
  {\bibinfo  {journal} {Phys. Rev. Lett.}\ }\textbf {\bibinfo {volume} {104}},\
  \bibinfo {pages} {240401} (\bibinfo {year} {2010})}\BibitemShut {NoStop}%
\bibitem [{\citenamefont {Dixon}\ \emph {et~al.}(2009)\citenamefont {Dixon},
  \citenamefont {Starling}, \citenamefont {Jordan},\ and\ \citenamefont
  {Howell}}]{Dixon2009}%
  \BibitemOpen
  \bibfield  {author} {\bibinfo {author} {\bibfnamefont {P.~B.}\ \bibnamefont
  {Dixon}}, \bibinfo {author} {\bibfnamefont {D.~J.}\ \bibnamefont {Starling}},
  \bibinfo {author} {\bibfnamefont {A.~N.}\ \bibnamefont {Jordan}}, \ and\
  \bibinfo {author} {\bibfnamefont {J.~C.}\ \bibnamefont {Howell}},\ }\href
  {\doibase 10.1103/PhysRevLett.102.173601} {\bibfield  {journal} {\bibinfo
  {journal} {Phys. Rev. Lett.}\ }\textbf {\bibinfo {volume} {102}},\ \bibinfo
  {eid} {173601} (\bibinfo {year} {2009})}\BibitemShut {NoStop}%
\bibitem [{\citenamefont {Hori}\ \emph {et~al.}(1989)\citenamefont {Hori},
  \citenamefont {Matsui}, \citenamefont {Araki},\ and\ \citenamefont
  {Inomata}}]{Hori1989}%
  \BibitemOpen
  \bibfield  {author} {\bibinfo {author} {\bibfnamefont {T.}~\bibnamefont
  {Hori}}, \bibinfo {author} {\bibfnamefont {T.}~\bibnamefont {Matsui}},
  \bibinfo {author} {\bibfnamefont {K.}~\bibnamefont {Araki}}, \ and\ \bibinfo
  {author} {\bibfnamefont {H.}~\bibnamefont {Inomata}},\ }\href@noop {}
  {\bibfield  {journal} {\bibinfo  {journal} {Opt. Lett.}\ }\textbf {\bibinfo
  {volume} {14}},\ \bibinfo {pages} {302} (\bibinfo {year} {1989})}\BibitemShut
  {NoStop}%
\bibitem [{\citenamefont {Nishimiya}\ \emph {et~al.}(2004)\citenamefont
  {Nishimiya}, \citenamefont {Yamaguchi}, \citenamefont {Ohrui},\ and\
  \citenamefont {Suzuki}}]{Nishimiya2004}%
  \BibitemOpen
  \bibfield  {author} {\bibinfo {author} {\bibfnamefont {N.}~\bibnamefont
  {Nishimiya}}, \bibinfo {author} {\bibfnamefont {Y.}~\bibnamefont
  {Yamaguchi}}, \bibinfo {author} {\bibfnamefont {Y.}~\bibnamefont {Ohrui}}, \
  and\ \bibinfo {author} {\bibfnamefont {M.}~\bibnamefont {Suzuki}},\ }\href
  {\doibase DOI: 10.1016/S1386-1425(03)00222-1} {\bibfield  {journal} {\bibinfo
   {journal} {Spectrochim. Acta, Part A}\ }\textbf {\bibinfo {volume} {60}},\
  \bibinfo {pages} {493 } (\bibinfo {year} {2004})}\BibitemShut {NoStop}%
\bibitem [{\citenamefont {Notcutt}\ \emph {et~al.}(2005)\citenamefont
  {Notcutt}, \citenamefont {Ma}, \citenamefont {Ye},\ and\ \citenamefont
  {Hall}}]{Notcutt2005}%
  \BibitemOpen
  \bibfield  {author} {\bibinfo {author} {\bibfnamefont {M.}~\bibnamefont
  {Notcutt}}, \bibinfo {author} {\bibfnamefont {L.-S.}\ \bibnamefont {Ma}},
  \bibinfo {author} {\bibfnamefont {J.}~\bibnamefont {Ye}}, \ and\ \bibinfo
  {author} {\bibfnamefont {J.~L.}\ \bibnamefont {Hall}},\ }\href@noop {}
  {\bibfield  {journal} {\bibinfo  {journal} {Opt. Lett.}\ }\textbf {\bibinfo
  {volume} {30}},\ \bibinfo {pages} {1815} (\bibinfo {year}
  {2005})}\BibitemShut {NoStop}%
\bibitem [{\citenamefont {Tropea}(1995)}]{Tropea1995}%
  \BibitemOpen
  \bibfield  {author} {\bibinfo {author} {\bibfnamefont {C.}~\bibnamefont
  {Tropea}},\ }\href@noop {} {\bibfield  {journal} {\bibinfo  {journal} {Meas.
  Sci. Technol.}\ }\textbf {\bibinfo {volume} {6}},\ \bibinfo {pages} {605}
  (\bibinfo {year} {1995})}\BibitemShut {NoStop}%
\bibitem [{\citenamefont {Pohl}\ \emph {et~al.}(2010)\citenamefont {Pohl} \emph
  {et~al.}}]{Pohl2010}%
  \BibitemOpen
  \bibfield  {author} {\bibinfo {author} {\bibfnamefont {R.}~\bibnamefont
  {Pohl}} \emph {et~al.},\ }\href {\doibase 10.1038/nature09250} {\bibfield
  {journal} {\bibinfo  {journal} {Nature}\ }\textbf {\bibinfo {volume} {466}},\
  \bibinfo {pages} {213} (\bibinfo {year} {2010})}\BibitemShut {NoStop}%
\bibitem [{\citenamefont {Ritchie}\ \emph {et~al.}(1991)\citenamefont
  {Ritchie}, \citenamefont {Story},\ and\ \citenamefont {Hulet}}]{Ritchie1990}%
  \BibitemOpen
  \bibfield  {author} {\bibinfo {author} {\bibfnamefont {N.~W.~M.}\
  \bibnamefont {Ritchie}}, \bibinfo {author} {\bibfnamefont {J.~G.}\
  \bibnamefont {Story}}, \ and\ \bibinfo {author} {\bibfnamefont {R.~G.}\
  \bibnamefont {Hulet}},\ }\href@noop {} {\bibfield  {journal} {\bibinfo
  {journal} {Phys. Rev. Lett.}\ }\textbf {\bibinfo {volume} {66}},\ \bibinfo
  {pages} {1107} (\bibinfo {year} {1991})}\BibitemShut {NoStop}%
\bibitem [{\citenamefont {Pryde}\ \emph {et~al.}(2005)\citenamefont {Pryde},
  \citenamefont {O'Brien}, \citenamefont {White}, \citenamefont {Ralph},\ and\
  \citenamefont {Wiseman}}]{Pryde2005}%
  \BibitemOpen
  \bibfield  {author} {\bibinfo {author} {\bibfnamefont {G.~J.}\ \bibnamefont
  {Pryde}}, \bibinfo {author} {\bibfnamefont {J.~L.}\ \bibnamefont {O'Brien}},
  \bibinfo {author} {\bibfnamefont {A.~G.}\ \bibnamefont {White}}, \bibinfo
  {author} {\bibfnamefont {T.~C.}\ \bibnamefont {Ralph}}, \ and\ \bibinfo
  {author} {\bibfnamefont {H.~M.}\ \bibnamefont {Wiseman}},\ }\href {\doibase
  10.1103/PhysRevLett.94.220405} {\bibfield  {journal} {\bibinfo  {journal}
  {Phys. Rev. Lett.}\ }\textbf {\bibinfo {volume} {94}},\ \bibinfo {pages}
  {220405} (\bibinfo {year} {2005})}\BibitemShut {NoStop}%
\bibitem [{\citenamefont {Starling}\ \emph {et~al.}(2010)\citenamefont
  {Starling}, \citenamefont {Dixon}, \citenamefont {Williams}, \citenamefont
  {Jordan},\ and\ \citenamefont {Howell}}]{Starling2010}%
  \BibitemOpen
  \bibfield  {author} {\bibinfo {author} {\bibfnamefont {D.~J.}\ \bibnamefont
  {Starling}}, \bibinfo {author} {\bibfnamefont {P.~B.}\ \bibnamefont {Dixon}},
  \bibinfo {author} {\bibfnamefont {N.~S.}\ \bibnamefont {Williams}}, \bibinfo
  {author} {\bibfnamefont {A.~N.}\ \bibnamefont {Jordan}}, \ and\ \bibinfo
  {author} {\bibfnamefont {J.~C.}\ \bibnamefont {Howell}},\ }\href {\doibase
  10.1103/PhysRevA.82.011802} {\bibfield  {journal} {\bibinfo  {journal} {Phys.
  Rev. A}\ }\textbf {\bibinfo {volume} {82}},\ \bibinfo {pages} {011802}
  (\bibinfo {year} {2010})}\BibitemShut {NoStop}%
\bibitem [{\citenamefont {Aharonov}\ and\ \citenamefont
  {Vaidman}(1991)}]{Aharonov1991}%
  \BibitemOpen
  \bibfield  {author} {\bibinfo {author} {\bibfnamefont {Y.}~\bibnamefont
  {Aharonov}}\ and\ \bibinfo {author} {\bibfnamefont {L.}~\bibnamefont
  {Vaidman}},\ }\href@noop {} {\bibfield  {journal} {\bibinfo  {journal} {J.
  Phys. A: Math. Gen.}\ }\textbf {\bibinfo {volume} {24}},\ \bibinfo {pages}
  {2315} (\bibinfo {year} {1991})}\BibitemShut {NoStop}%
\bibitem [{\citenamefont {Steinberg}(2010)}]{Steinberg2010}%
  \BibitemOpen
  \bibfield  {author} {\bibinfo {author} {\bibfnamefont {A.~M.}\ \bibnamefont
  {Steinberg}},\ }\href {\doibase 10.1038/463890a} {\bibfield  {journal}
  {\bibinfo  {journal} {Nature}\ }\textbf {\bibinfo {volume} {463}},\ \bibinfo
  {pages} {890} (\bibinfo {year} {2010})}\BibitemShut {NoStop}%
\bibitem [{\citenamefont {Hardy}(1992)}]{Hardy1992}%
  \BibitemOpen
  \bibfield  {author} {\bibinfo {author} {\bibfnamefont {L.}~\bibnamefont
  {Hardy}},\ }\href {\doibase 10.1103/PhysRevLett.68.2981} {\bibfield
  {journal} {\bibinfo  {journal} {Phys. Rev. Lett.}\ }\textbf {\bibinfo
  {volume} {68}},\ \bibinfo {pages} {2981} (\bibinfo {year}
  {1992})}\BibitemShut {NoStop}%
\bibitem [{\citenamefont {Lundeen}\ and\ \citenamefont
  {Steinberg}(2009)}]{Lundeen2009}%
  \BibitemOpen
  \bibfield  {author} {\bibinfo {author} {\bibfnamefont {J.~S.}\ \bibnamefont
  {Lundeen}}\ and\ \bibinfo {author} {\bibfnamefont {A.~M.}\ \bibnamefont
  {Steinberg}},\ }\href {\doibase 10.1103/PhysRevLett.102.020404} {\bibfield
  {journal} {\bibinfo  {journal} {Phys. Rev. Lett.}\ }\textbf {\bibinfo
  {volume} {102}},\ \bibinfo {pages} {020404} (\bibinfo {year}
  {2009})}\BibitemShut {NoStop}%
\bibitem [{\citenamefont {Mitchison}\ \emph {et~al.}(2007)\citenamefont
  {Mitchison}, \citenamefont {Jozsa},\ and\ \citenamefont
  {Popescu}}]{Mitchison2007}%
  \BibitemOpen
  \bibfield  {author} {\bibinfo {author} {\bibfnamefont {G.}~\bibnamefont
  {Mitchison}}, \bibinfo {author} {\bibfnamefont {R.}~\bibnamefont {Jozsa}}, \
  and\ \bibinfo {author} {\bibfnamefont {S.}~\bibnamefont {Popescu}},\ }\href
  {\doibase 10.1103/PhysRevA.76.062105} {\bibfield  {journal} {\bibinfo
  {journal} {Phys. Rev. A}\ }\textbf {\bibinfo {volume} {76}},\ \bibinfo
  {pages} {062105} (\bibinfo {year} {2007})}\BibitemShut {NoStop}%
\bibitem [{\citenamefont {Hosten}\ and\ \citenamefont
  {Kwiat}(2008)}]{Hosten2008}%
  \BibitemOpen
  \bibfield  {author} {\bibinfo {author} {\bibfnamefont {O.}~\bibnamefont
  {Hosten}}\ and\ \bibinfo {author} {\bibfnamefont {P.}~\bibnamefont {Kwiat}},\
  }\href {\doibase 10.1126/science.1152697} {\bibfield  {journal} {\bibinfo
  {journal} {Science}\ }\textbf {\bibinfo {volume} {319}},\ \bibinfo {pages}
  {787} (\bibinfo {year} {2008})}\BibitemShut {NoStop}%
\bibitem [{\citenamefont {Starling}\ \emph {et~al.}(2009)\citenamefont
  {Starling}, \citenamefont {Dixon}, \citenamefont {Jordan},\ and\
  \citenamefont {Howell}}]{Starling2009}%
  \BibitemOpen
  \bibfield  {author} {\bibinfo {author} {\bibfnamefont {D.~J.}\ \bibnamefont
  {Starling}}, \bibinfo {author} {\bibfnamefont {P.~B.}\ \bibnamefont {Dixon}},
  \bibinfo {author} {\bibfnamefont {A.~N.}\ \bibnamefont {Jordan}}, \ and\
  \bibinfo {author} {\bibfnamefont {J.~C.}\ \bibnamefont {Howell}},\ }\href
  {\doibase 10.1103/PhysRevA.80.041803} {\bibfield  {journal} {\bibinfo
  {journal} {Phys. Rev. A}\ }\textbf {\bibinfo {volume} {80}},\ \bibinfo {eid}
  {041803 (R)} (\bibinfo {year} {2009})}\BibitemShut {NoStop}%
\bibitem [{\citenamefont {Williams}\ and\ \citenamefont
  {Jordan}(2008)}]{Williams2008}%
  \BibitemOpen
  \bibfield  {author} {\bibinfo {author} {\bibfnamefont {N.~S.}\ \bibnamefont
  {Williams}}\ and\ \bibinfo {author} {\bibfnamefont {A.~N.}\ \bibnamefont
  {Jordan}},\ }\href@noop {} {\bibfield  {journal} {\bibinfo  {journal} {Phys.
  Rev. Lett.}\ }\textbf {\bibinfo {volume} {100}},\ \bibinfo {eid} {026804}
  (\bibinfo {year} {2008})}\BibitemShut {NoStop}%
\bibitem [{\citenamefont {Howell}\ \emph {et~al.}(2010)\citenamefont {Howell},
  \citenamefont {Starling}, \citenamefont {Dixon}, \citenamefont {Vudyasetu},\
  and\ \citenamefont {Jordan}}]{Howell2010}%
  \BibitemOpen
  \bibfield  {author} {\bibinfo {author} {\bibfnamefont {J.~C.}\ \bibnamefont
  {Howell}}, \bibinfo {author} {\bibfnamefont {D.~J.}\ \bibnamefont
  {Starling}}, \bibinfo {author} {\bibfnamefont {P.~B.}\ \bibnamefont {Dixon}},
  \bibinfo {author} {\bibfnamefont {P.~K.}\ \bibnamefont {Vudyasetu}}, \ and\
  \bibinfo {author} {\bibfnamefont {A.~N.}\ \bibnamefont {Jordan}},\ }\href
  {\doibase 10.1103/PhysRevA.81.033813} {\bibfield  {journal} {\bibinfo
  {journal} {Phys. Rev. A}\ }\textbf {\bibinfo {volume} {81}},\ \bibinfo
  {pages} {033813} (\bibinfo {year} {2010})}\BibitemShut {NoStop}%
\bibitem [{\citenamefont {Brunner}\ and\ \citenamefont
  {Simon}(2010)}]{Brunner2010}%
  \BibitemOpen
  \bibfield  {author} {\bibinfo {author} {\bibfnamefont {N.}~\bibnamefont
  {Brunner}}\ and\ \bibinfo {author} {\bibfnamefont {C.}~\bibnamefont
  {Simon}},\ }\href {\doibase 10.1103/PhysRevLett.105.010405} {\bibfield
  {journal} {\bibinfo  {journal} {Phys. Rev. Lett.}\ }\textbf {\bibinfo
  {volume} {105}},\ \bibinfo {pages} {010405} (\bibinfo {year}
  {2010})}\BibitemShut {NoStop}%
\bibitem [{\citenamefont {Pedrotti}\ and\ \citenamefont
  {Pedrotti}(1993)}]{Pedrotti1993}%
  \BibitemOpen
  \bibfield  {author} {\bibinfo {author} {\bibfnamefont {F.~L.}\ \bibnamefont
  {Pedrotti}}\ and\ \bibinfo {author} {\bibfnamefont {L.~S.}\ \bibnamefont
  {Pedrotti}},\ }\href@noop {} {\emph {\bibinfo {title} {Introduction to
  Optics}}},\ \bibinfo {edition} {2nd}\ ed.\ (\bibinfo  {publisher}
  {Prentice-Hall, Inc.},\ \bibinfo {year} {1993})\BibitemShut {NoStop}%
\bibitem [{\citenamefont {Malitson}(1965)}]{Malitson1965}%
  \BibitemOpen
  \bibfield  {author} {\bibinfo {author} {\bibfnamefont {I.~H.}\ \bibnamefont
  {Malitson}},\ }\href@noop {} {\bibfield  {journal} {\bibinfo  {journal} {J.
  Opt. Soc. Am.}\ }\textbf {\bibinfo {volume} {55}},\ \bibinfo {pages} {1205}
  (\bibinfo {year} {1965})}\BibitemShut {NoStop}%
\bibitem [{\citenamefont {Steck}(2010)}]{Steck2010}%
  \BibitemOpen
  \bibfield  {author} {\bibinfo {author} {\bibfnamefont {D.~A.}\ \bibnamefont
  {Steck}},\ }\emph {http://steck.us/alkalidata/} {\  (\bibinfo {year}
  {2010})}\BibitemShut {NoStop}%
\end{thebibliography}
\end{document}